\documentclass[pra,aps,nofootinbib,showkeys,showpacs,twocolumn]{revtex4}
\usepackage{epsfig}
\usepackage{graphicx}
\usepackage{amssymb}
\usepackage{color}
\usepackage{amsmath}

\begin{document}

\title{From resonantly interacting fermions with effective range to neutron matter
%A Local density functional for Fermi gas: from unitary to non-unitary regime including effective range correction 
% DFT for resonantly interacting Fermi systems including effective range effect
}
  
\author{Denis Lacroix} \email{lacroix@ipno.in2p3.fr}
\affiliation{Institut de Physique Nucl\'eaire, IN2P3-CNRS, Universit\'e Paris-Sud, Universit\'e Paris-Saclay, F-91406 Orsay Cedex, France}
\date{\today}
\begin{abstract} 
%Inspired by the Effective Field Theory (EFT) recent progress, 
A density functional theory is proposed for strongly 
interacting fermions with arbitrary large negative scattering length. The functional has only two parameters that are directly fixed to reproduce 
the universal properties of unitary gas: the so-called "Bertsch parameter" $\xi_0$ and a parameter $\eta_e$ related to the possible 
influence of the effective range $r_e$ at infinite scattering length $a$. Using most recent quantum Monte-Carlo (QMC) estimates 
of these two parameters, it is shown that the functional properly reproduces the experimental measurements of interacting Fermi systems 
not only at unitarity but also away from this limit over a wide range of $(ak_F)^{-1}$ values. The functional is applied to obtain an expression of the Tan's 
contact parameter including the effect of $r_e$. 
%Good agreement with various experimental measurements is obtained. 
Application is finally made to neutron matter. It is shown that most recent QMC results are well reproduced. 
\end{abstract}

\pacs{67.85.Lm,21.65.-f}
  
\keywords{strongly interacting fermions, unitary limit, neutron matter}

\maketitle

During the last ten years, important progresses have been made to 
manipulate Fermi gas by tuning the interaction between particles \cite{Blo08,Chi10}. 
A special attention has been paid close to the unitarity regime when the s-wave scattering length   
becomes infinite in dilute systems. In this case, the interacting system properties become universal.
The energy of the system is directly proportional to the free Fermi gas energy, and the ratio 
between these two energies is a universal parameter $\xi_0 \simeq 0.37$ \cite{Gan11,End11,Ku12,Zur13}, the so-called "Bertsch parameter" \cite{Ber00}. 
The experimental advances in atomic Fermi systems have motivated tremendous theoretical efforts to understand systems at unitarity as well as the transition from BCS to BEC regimes (see for instance the collection of review articles in 
Ref. \cite{Zwe11}).  Due to the universal behavior of the energy for unitary gases, it was shown that simple local density functionals 
directly adjusted on QMC approaches can accurately describe various static or dynamical properties of these systems
\cite{Pap05,Pap06,Bul07,Bul12}. These functionals however strictly apply for $|a| \rightarrow \pm \infty$ in the low density regime and 
cannot describe unitary gases with possibly non-zero effective range $r_e$. The description Fermi gas with non-zero effective range and anomalously large $a$
is motivated by (i) the possibility to uncover new effects in a wider class of unitary systems (ii) neutron systems that enters into such class 
of interacting fermions. 

A first step to include effective range influence is 
made in Refs \cite{For11,For12} showing non-trivial effects due to $r_e$. In this work, a minimal generalization of the previously proposed DFT is made 
by allowing explicitly the parameters of the functional to vary with $r_e$.   Alternatively, finite--range effects can be investigated 
at low--density using Effective Field Theory (EFT) and systematic expansion in the Fermi momentum (see for instance \cite{Ham00,Fur12}). One success of EFT is 
the improvement of the universal Lee-Yang (LY) formula \cite{Lee57,Bis73}, including $r_e$ effects for low density Fermi systems.  
The EFT approach however cannot be directly applied to unitary gases unless specific resummations of infinite order in-medium loops are made. This has led to the description of systems at or close to unitary with varying success \cite{Ste00,Sch05,Kai11}. Here, inspired by the EFT approach with resummation, a new DFT for Fermi systems, that 
is optimized at unitarity and smoothly behaves away from unitarity, is proposed .

I consider here spin degenerated Fermionic systems with a s-wave interaction characterized by a negative scattering length $a$ and an effective range $r_e$. These two quantities are defined as usual as the leading and next to leading order of the 
expansion of the s-wave phase-shift $\delta$ in terms relative momentum $k$ of the interacting particles:
\begin{eqnarray}
k \cot \delta &=& - \frac{1}{a} + \frac{1}{2} r_e k^2 + O(k^4).
\end{eqnarray}

\begin{figure}[htbp]
\includegraphics[width=\linewidth]{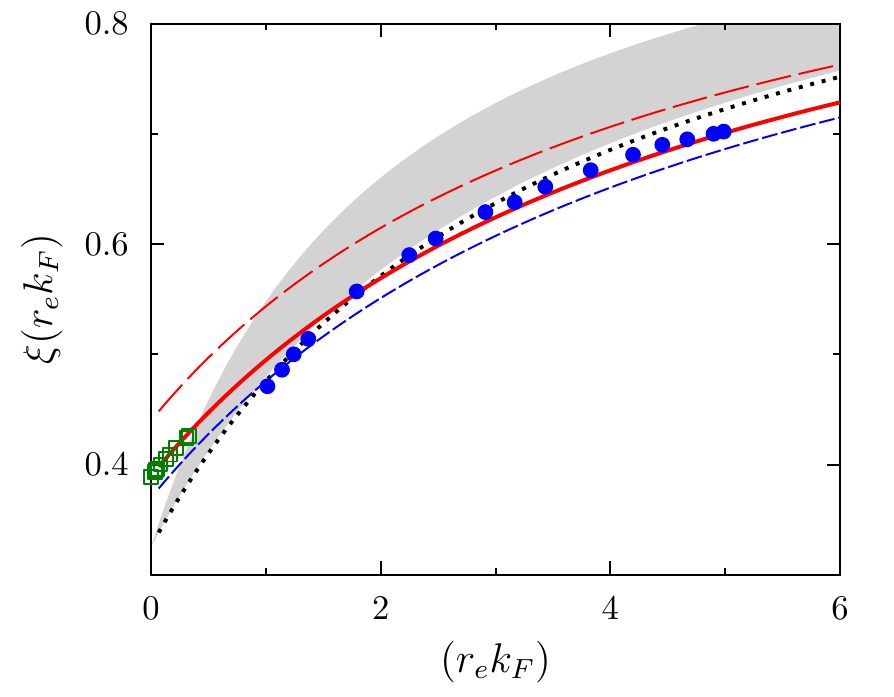}
\vspace{-5mm}
\caption{ (color online) Illustration of the formula (\ref{eq:rskf1}) (black dotted curve) giving the effective range dependence of the $\xi$ parameter for unitary gas. The results are 
compared to the one of Ref. \cite{Sch05_2} (blue point)
and to the results at low density of   \cite{For11,For12} (green open squares). 
Note that in \cite{Sch05_2}, results are given for the Hartree-Fock energy only. The red solid line 
is obtained using Eq. (\ref{eq:unitaryopt}) with the values given in ref. \cite{For11,For12}: $\xi_0 =0.3897$ and $\eta_e =0.127$. The short--dashed blue line and 
long--dashed red line correspond
to $\xi_0 = 0.37$ and $\xi_0=0.44$ respectively both with  $\eta_e =0.127$.
These values are compatible with those reported in ref. \cite{Gan11}.
The grey area corresponds to the results given in ref. \cite{Sch05} and reflects the renormalization scale dependence in EFT calculations. 
The latter case includes resummation of correlation effects.}
\label{fig:unit1} 
\end{figure}

The functional proposed below is strongly guided by the resummation technique used 
in EFT to tackle the problem of unitary gas. In this case, simplified resummed formula 
have been obtained in \cite{Ste00,Sch05,Kai11}. 
Here, I introduce a resummed formula that account for non-zero $r_e$
and write the energy as a functional of the Fermi momentum $k_F$ as follows:
\begin{eqnarray}
\frac{E}{N}=\frac{\hbar^2k_F^2}{2m}\left\{
\frac{3}{5}+  \frac{(a k_F) A_0  }{1 - A_0^{-1} \left[ A_1 + (r_e k_F)A_2   \right] a k_F  } \right\}. \label{eq:resumgen}
\end{eqnarray}
$(A_0,A_1,A_2)$ are three constants to be determined.  One possibility that has been explored in \cite{Sch05,Yan15}, neglecting possible $r_e$ effects, is 
to constrain the functional by matching the low density limit with the universal Lee-Yang expansion of the energy \cite{Lee57,Bis73}.  This gives $A_0=2/(3 \pi)$,  
\begin{eqnarray}
A_1= \frac{4}{35\pi^2} (11-2 \ln2)~{\rm and}~A_2 =  \frac{1}{10 \pi}. \nonumber
\end{eqnarray}
The advantages of the formula (\ref{eq:resumgen}) are that (i) it can be reinterpreted as a density functional theory through the relation 
$k_F = (3 \pi^2 n)^{1/3}$ where $n$ is the density (ii) it can be expanded in powers of
$(ak_F)$, $(a k_F)^{-1}$ or $(r_e k_F)$; (iii) it has a definite limit as $a \rightarrow + \infty$. Taking this specific limit, we obtain ${E} = \xi(r_ek_F) E_{\rm FG}$ where $E_{\rm FG}$ is the free Fermi gas energy, $E_{\rm FG}/N=\left[3 \hbar^2 k_F^2/(10m)\right]$, and
\begin{eqnarray}
\xi(r_ek_F) & = & \left\{1 -  \frac{5}{3} \frac{A^2_0}{\left( A_1 + (r_e k_F )  A_2 \right)} \right\}. \label{eq:rskf1}
\end{eqnarray}  
For unitary gas, with small but non zero effective range, this expression can be expanded as:
\begin{eqnarray}
\xi(r_ek_F)& \simeq & \left( 1 -  \frac{5}{3} \frac{A^2_0}{A_1} \right)  + \frac{5}{3}\frac{A^2_0 A_2}{A^2_1} ( r_e k_F)  - \frac{5}{3}\frac{A^2_0 A^2_2}{A^3_1} ( r_e k_F)^2  \nonumber \\
&\equiv &\xi_0 + ( r_e k_F) \eta_e   + ( r_e k_F)^2\delta_e  .\label{eq:etae}
%\left\{1 -  \frac{5}{3} \frac{A^2_0}{\left( A_1 + A_2 (k_F r_e) \right)} \right\}. \label{eq:rskf1}
\end{eqnarray}
With values of $\{ A_i \}$ parameters deduced from low density constraint, one get $\xi_0 = 0.326$, $\eta_e=0.19$ and $\delta_e=-0.055$. The two former values have been obtained 
using a phase-space average in ref. \cite{Sch05}. Not only $\xi_0$ is close to the expected value for unitary gas, i.e. 
$\xi_0\simeq 0.37$, but also, 
the effective range dependence is not too far from the result obtained with fixed-node QMC calculations of Ref. \cite{For11,For12}. More generally the $(r_e k_F)$ dependence deduced from Eq. (\ref{eq:rskf1}), while not perfect, is globally consistent with elaborated T-matrix estimates \cite{Sch05_2,Sch05} (see Fig. \ref{fig:unit1}).   
The Lee-Yang formula contains solely mean-field and second order perturbation term.  
The fact that a constraint on LY is close to the observed properties at unitarity indicates that strong 
cancellations of higher-order effects occur.  
Attempts have been made to further improve the expression of the energy by resumming higher order effects \cite{Ste00,Sch05,Kai11}, leading to various results for $\xi_0$. However, the values obtains strongly depend on the strategy to select loops in the expansion.

\begin{figure}[htbp]
\includegraphics[width=\linewidth]{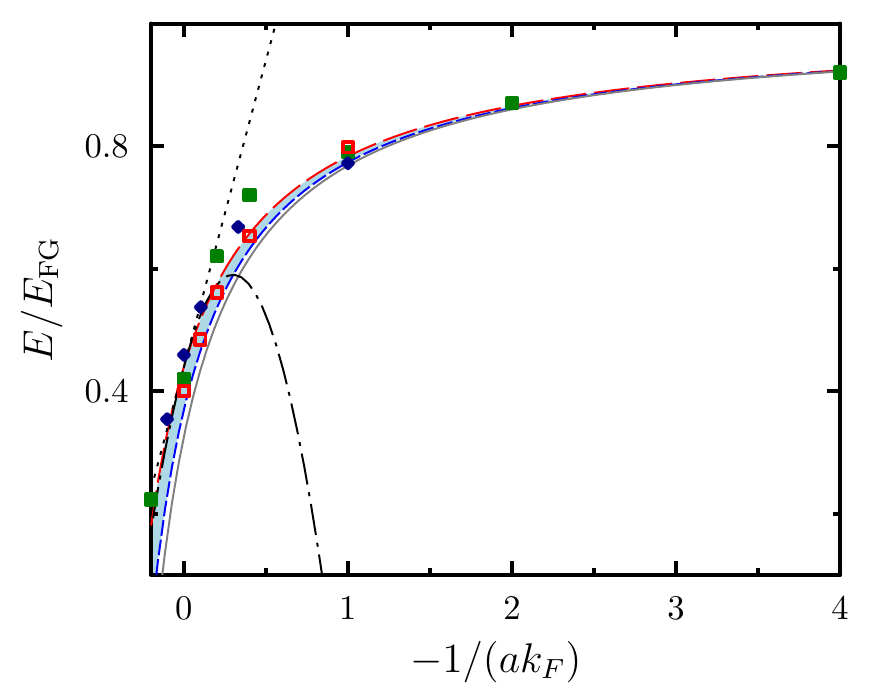}
\vspace{-5mm}
\caption{ (color online) Energy of a Fermi gas with zero effective range obtained with the Monte-Carlo calculations \cite{Cha04} (dark blue diamond), \cite{Ast04} (green square) and the more recent result of Ref. \cite{Gez08} (red open square).
The results of formula (\ref{eq:start1}) with $\xi( r_e k_F) = \xi_0$ are shown for $\xi_0=0.37$ (blue short dashed line) 
and $\xi_0=0.44$ (red long dashed line). The blue area illustrates the uncertainty on $\xi_0$.
For comparison, the first and second order expansion in $(ak_F)^{-1}$, Eq. (\ref{eq:exp2}), using $\zeta=\nu=1$ (see discussion in \cite{Bul05}) are displayed respectively with black dotted and black dot-dashed lines.  Results obtained using directly the functional constrained with the Lee-Yang expansion ($\xi_0=0.326$) together with Eq. (\ref{eq:start1}) are also displayed with grey solid line. 
}
\label{fig:unit2} 
\end{figure}

As a intermediate summary, the functional (\ref{eq:resumgen}) has many interesting features. However, starting from the low density expansion, it only provides a very rough description of Fermi gas at and around unitarity. 

For this reason, here I follow a more pragmatic strategy and directly constrain the functional (\ref{eq:resumgen}) assuming that its Taylor expansion match Eq. (\ref{eq:etae}) up to first order in $( r_e k_F)$. Then, the functional will depend only on the two parameters $\xi_0$ and $\eta_e$.  Since (\ref{eq:resumgen}) has 3 parameters and the Taylor expansion 
around unitarity only leads to 2 constraints, I again assume that $A_0=2/(3 \pi)$ so that the LY formula is also reproduced up to $k^3_F$ in dilute systems.  The resulting energy then writes:
\begin{eqnarray}
\frac{E}{N}  =  \frac{\hbar^2k_F^2}{2m} \left\{ \frac{3}{5} + \frac{2}{3 \pi} \frac{ (a k_F)}{1 - \frac{10}{9 \pi} (a k_F)/ (1 - \xi( r_e k_F))} \right\},\label{eq:start1}
\end{eqnarray} 
with:
\begin{eqnarray}
\xi( r_e k_F)  = \left\{ 1 - \frac{(1-\xi_0)^2}{(1-\xi_0) + ( r_e k_F)\eta_e } \right\}. \label{eq:unitaryopt} 
\end{eqnarray}
These two formulas are the main results of the present letter. In the following, I explore the range of applicability 
of the novel functional as well as possible applications. Note that $\xi_0$ and $\eta_e$ are not free parameters in the sense that they are fixed 
by recent measurements or QMC calculations in unitary gas. 

In Fig. \ref{fig:unit2}, an illustration of the energy obtained as a function of $(a k_F)^{-1}$ for different $\xi_0$ (and $r_e=0$) are shown and compared to experiments.  It is interesting to mention that the result obtained using directly the Lee-Yang expansion as a constraint, i.e.
using Eq. (\ref{eq:start1}) with $\xi_0 = 0.326$, gives already a good approximation especially for $-(a k_F)^{-1} > 1$. However, around unitarity, the energy is underestimated. With constraints imposed at unitarity, the energy is reasonably reproduced both as $1/(a k_F) \rightarrow 0$ and for large values of $(a k_F)^{-1}$.

\begin{figure}[htbp]
\includegraphics[width=\linewidth]{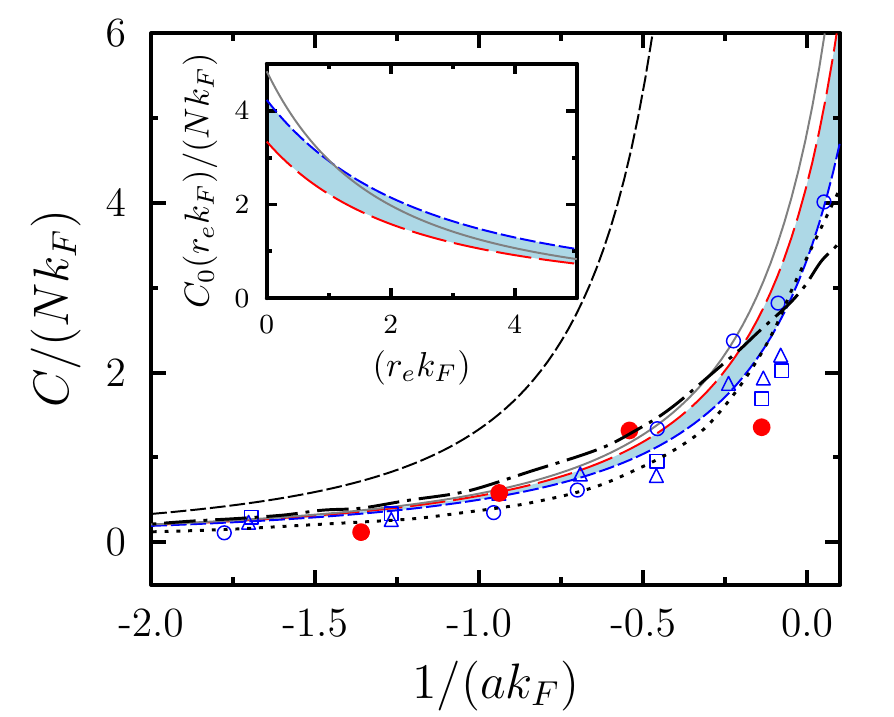}
\vspace{-5mm}
\caption{ (color online) Contact parameter as a function of $(a k_F)^{-1}$ obtained with Eq.  (\ref{eq:CN}) assuming $r_e=0$. The light blue area 
is the region between the two results obtained either with $\xi_0=0.37$ and $\xi_0 = 0.44$. The black dashed line is the BCS result given by $4 (a k_F)^2/3$.
The black dotted and dot-dashed lines are the theoretical results of Ref. \cite{Wer09} and Ref. \cite{Hau09} respectively. The blue open circles, triangles and 
squares are the moment, PES and rf measurements of \cite{Ste10}. The red filled circles is the measure of Ref. \cite{Kuh10}. 
In the inset, 
 the dependence of the contact parameter for unitarity gas with non-zero effective range $r_e$ is shown
(for $\xi_0$ from $0.37$ to $0.44$) using a value $\eta_e=0.127$ in Eq. (\ref{eq:CN0}). In the figure and in the inset, the gray solid line corresponds to the contact parameter obtained using the values $\xi_0 = 0.326$ and $\eta_e = 0.19$ (constraint with the low density Lee-Yang expansion).}
\label{fig:C} 
\end{figure} 

It is worth noticing that, for $r_e=0$, close to unitarity we deduce from Eq. (\ref{eq:start1}):
 \begin{eqnarray}
\frac{E}{N}  &\simeq& \frac{3}{5}  \frac{\hbar^2k_F^2}{2m} \left\{\xi_0  -  \frac{\zeta}{(a k_F)} - \frac{5}{3} \frac{\nu}{(a k_F)^2} + \cdots
 \right\} \label{eq:exp2}
 %&\simeq& \frac{3}{5}  \frac{\hbar^2k_F^2}{2m} \left\{\xi_0  - \frac{9 \pi}{10} \frac{(1-\xi_0)^2}{(k_F a)} 
 %- \left(\frac{9 \pi}{10} \right)^2  \frac{(1-\xi_0)^3}{(k_F a)^2}  + \cdots  \right\} \nonumber \\
% &\equiv&
  \end{eqnarray}
with 
\begin{eqnarray}
\zeta & = & \frac{9 \pi}{10} (1-\xi_0)^2~~{\rm and}~~ \nu = \frac{3}{5} \frac{\zeta^2}{(1-\xi_0)}
\end{eqnarray} 
For $\xi_0=0.37$, this gives $\zeta=1.12$ and $\nu=1.19$, while for $\xi_0=0.44$, we obtain $\zeta=0.89$ and $\nu=0.84$.  These values 
are rather close from those reported in QMC ($\zeta =0.9$) \cite{Cha04,Ast04} or fixed node diffusion Monte-Carlo ($\zeta =0.95$) \cite{Lob06}.
In Ref. \cite{Bul05}, a direct fit to QMC leads also to $\zeta \simeq \nu \simeq 1$. The range of validity of the expansion (\ref{eq:exp2}) is illustrated 
by the dotted and dot-dashed lines in Fig. \ref{fig:unit2}. We clearly see that it applies only for $1/|ak_F| \ll 1$ while the new functional 
applies on a wider range of density.
%r_e depe
Using directly $\xi_0$ and $\eta_e$ to constraint the functional also leads to a better description of the possible effective range dependence. In Fig. \ref{fig:unit1}, I present the quantity $\xi ( r_e k_F) $, given by Eq. (\ref{eq:unitaryopt}) using the values $\xi_0 =0.3897$ and $\eta_e =0.127$ \cite{For11,For12}. The QMC result with non zero effective range are perfectly  described while keeping a description at larger $( r_e k_F)$ consistent with the many-body calculations of Refs. \cite{Sch05,Sch05_2}.

I now illustrate how the functional can be used, first to reproduce some observables measured in unitary gas and second, to predict their 
possible effective range dependence. A typical example, that has been the subject of intensive experimental and theoretical works \cite{Bra11}
is the Tan's contact parameter \cite{Tan08_1,Tan08_2,Tan08_3}. Here, I follow closely \cite{Bra11}. 
In the present case of infinite spin saturated system, the contact parameter $C$ can be related to the contact density ${\cal C}$ 
through $C / (N k_F) =  (3 \pi^2) {\cal C}/{k_F^4}$ that is itself related to the energy density ${\cal E}$ through  ${\cal C}={4\pi m a^2} \left(d{\cal E} /d a \right)/{\hbar^2}$. The energy density is given by ${\cal E}= k^3_F E/(3 \pi^2 N)$. Using the expression (\ref{eq:start1}), we deduce that the contact parameter 
expresses as:
\begin{eqnarray}
\frac{C}{Nk_F} & = &  \frac{4}{3}\frac{(a k_F)^2}{\left[1 - \frac{10}{9 \pi} (a k_F)/ (1 - \xi( r_e k_F)) \right]^2}. \label{eq:CN}
\end{eqnarray}
The resulting contact parameter is shown as a function of $(ak_F)^{-1}$ in figure \ref{fig:C} for the specific case 
$r_e=0$. $C$ deduced from the new functional is in good agreement with the experimental observations and within 
the errorbars of most recent theoretical estimates. We also show for comparison the result of the functional obtained using parameters 
deduced from the low density Lee-Yang expansion. It is interesting to note that these results are very close to the one obtained by setting
the constraint at unitarity. In particular, different curves cannot be distinguished for $(ak_F)^{-1} < -1$. However, around unitarity, differences are noticeable. It should also be mentioned that the comparison between Eq. (\ref{eq:CN}) and experiments is given here only as an illustration since this formula applies to uniform systems while experimental data are performed with finite-size non-uniform atomic clouds. A consistent comparison would require performing calculations accounting for finite-size effects using for example a local density approximation. This development will be considered in the near future.

Focusing now on the unitarity limit, I denote by $C_0$ the contact parameter as $a\rightarrow + \infty$. We then have:
\begin{eqnarray}
\frac{C_0}{N k_F}& = & \frac{27}{25}  \pi^2(1 - \xi(r_e k_F))^2, \label{eq:CN0}
\end{eqnarray} 
 that includes possible effective range dependence.  This dependence is illustrated in the inset of Fig. \ref{fig:C}.  It is predicted 
 that a reduction of $C_0$ should occur for unitary gas with non-zero effective range. The experimental observation of such reduction would be a stringent test of the present functional theory. In addition, since the contact parameter in the unitary limit has direct connection with many properties \cite{Bra11}, number of closed channel, shear viscosity, tail of the momentum distribution, I also anticipate that these 
 properties will be modified as $r_e$ increases while keeping the scattering length infinite. 
 Note however that most connection with the Tan parameter have been
 obtained using the zero-range model and a more careful study for non-zero $r_e$ is desirable. It is finally interesting to mention that the present functional can be directly combined with the technique proposed 
 in Ref. \cite{Wer09} to obtain the contact parameter for finite systems in a trap
 by re-interpreting the energy  as  a functional of the local density $n({\mathbf r})$. 

An realistic example of system that has large but finite negative scattering length and non-zero effective range is the case of 
neutron matter.  In this case, the neutron-neutron scattering length is $a \simeq -18.9$ fm and $r_e\simeq 2.7$ fm. In particular, the ratio
$|r_e/a| \simeq 0.14$ is much larger than the one generally obtained in cold atoms (see for instance discussion in Ref. \cite{Sch05}).
\begin{figure}[htbp]
\includegraphics[width=\linewidth]{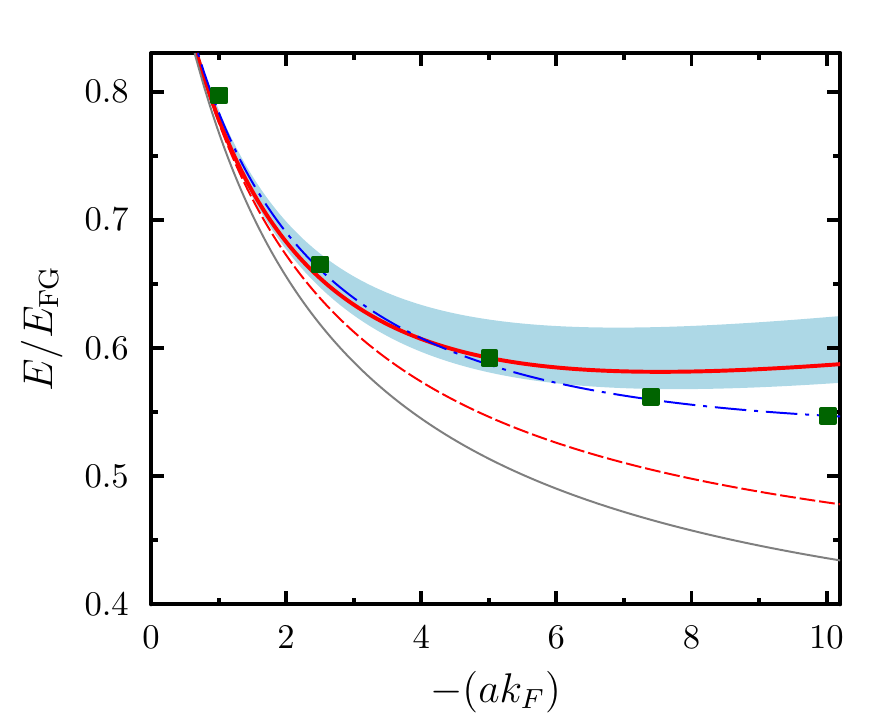}
\vspace{-5mm}
\caption{ (color online) Energy obtained with Quantum Monte-Carlo approach 
in neutron matter assuming a finite range s-wave interaction (green square) \cite{Gez10}.
The red curves are the result obtained with formula (\ref{eq:start1}) together with (\ref{eq:unitaryopt}) assuming $\xi_0 = 0.3897$, $\eta_e=0.127$ (thick solid line) \cite{For11,For12} or $\eta_e=0$ (dashed line). 
The blue area indicates the variation of the result when  $\xi_0$ varies from 0.37 to 0.44 keeping $\eta_e$ fixed. For comparison, I also show the result obtained 
using $\xi_0 = 0.44$ and $\eta_e=0.03$ that perfectly match the QMC approach (blue dot-dashed line). In all cases presented here, I took $a = -18.9$ fm and $r_e = 2.7$ fm. For comparison, I also show the result obtained (thin solid gray line) using the Lee-Yang expansion as a constraint ($\xi_0=0.326$ and $\eta_e=0.19$).}
\label{fig:unit4} 
\end{figure}
Consequently, density functional based on EFT techniques together with low density expansion only applies in a very narrow range of densities 
verifying $|a k_F| \ll 1$ and effective range effect should be included \cite{Ham00,Fur12}.  In Figure \ref{fig:unit4}, 
the  results of the new functional are compared with QMC calculations \cite{Gez10,Car12,Gan15}. For the comparison, I did not adjust 
the $\xi_0$ and $\eta_e$ parameters but directly took the values reported in Refs. \cite{For11,For12}. The blue area indicates the possible dependence of the result with the $\xi_0$ value. The new functional (red solid line) reproduces 
remarkably well the exact QMC results  even if $(a k_F)$ is much greater than 1. Some deviations are observed for $|ak_F| > 5$. 
This deviation might come from the fact that Eq. (\ref{eq:unitaryopt}) is constrained to reproduce the QMC only up to second order in $(r_e k_F)$ and higher order corrections might be needed. 
It is worth 
to mention that these deviations can also be reduced either by keeping $\xi_0=0.3897$ and reducing $\eta_e$ slightly to $0.08$ or by varying both parameters.
An example is given in Fig. \ref{fig:unit4} with dot-dashed line. The effect of $r_e$ is clearly pointed out by comparing the 
full results with the result obtained by keeping the same value of $\xi_0$ and setting artificially $\eta_e$ to 0. Important differences are observed 
uncovering the non-negligible effect of the effective range.  I note that the result obtained empirically in Ref. \cite{Ste00} 
is also close to the QMC. However, this result were obtained neglecting the effect of $r_e$ and the agreement is most probably accidental.  
In Ref. \cite{Yan15}, an alternative functional has been proposed and shown to match the equation of state of neutron matter from very
low to higher density. In this case, the parameters were adjusted to directly reproduce a set of ab-initio calculations. Restricting 
the range of density considered, it is shown that the new functional proposed here can reproduce exact QMC results by having as unique 
parameters the universal parameters deduced independently at unitarity. This clearly open new perspectives to less empirical nuclear  density functional theories.     

Another important conclusion is that the results obtained using parameters deduced only from low density constraints (gray thin solid line) 
fail to describe the neutron matter in the density range considered here. This indicates that for nuclear systems with very large scattering length in the s-wave channel and non-zero effective range, the unitary regime seems to be a better starting point than the low density expansion. Interestingly enough this strategy has been  recently explored in Ref. \cite{Kon16} where calculations are made for small nuclei setting the unitary limit as the leading order.

In the present work, inspired by the resummed expression obtained in EFT, I propose a new functional 
for strongly interacting Fermi systems. The functional parameters are directly constrained to reproduce the universal behavior 
of Fermi gas at unitarity including the possible effective range influence. The resulting functional has only two parameters, the 
so-called "Bertsch parameter", $\xi_0$ and the effective range parameter $\eta_e$ that can be taken from previous studies
on unitary gas. 
The proposed functional  further has the advantage to naturally  extend functionals \cite{Pap05,Pap06,Bul07,Bul12} proposed for unitary gas at low density while being able to (i) describe systems with finite scattering length, (ii)
include possible effective range effects and (iii) applies on a wider range of density, even if $|a k_F| \gg 1$.
I apply the functional to estimate the energy of Fermi gas on a wide range of $(ak_F)$ values showing good 
reproduction of experimental  results as well as QMC calculations. I use the functional to obtain an expression 
of the Tan's contact parameter at and away from unitarity.  Possible 
effect of non-zero effective range at unitarity are also discussed.  It is finally shown that the functional can reproduce well 
recent QMC results obtained for symmetric neutron matter. 

The present work promotes the idea that the unitary regime can be a proper starting point for describing systems with large 
s-wave scattering length and eventually non-zero effective range. Here, I concentrated on infinite uniform systems. To apply the 
novel functional to finite systems like atomic clouds and/or nuclei a number of issues should be clarified.  The most natural 
way to extend a functional designed for infinite systems to finite systems is to use a local density approximation. Then, the energy 
becomes functional of the local density. It is anticipated however, that such a rough approximation will improperly describe 
non-uniform systems. For instance, the effective mass of nuclear or atomic systems are known to differ from the bare mass. A 
proper treatment of the effective mass, would require to introduce explicit gradient correction in the functional.   

A second important aspect is superfluidity that is crucial in both nuclei and atomic clouds. The energy provided by the functional 
(\ref{eq:start1}) is describing the total energy of interacting fermionic systems and therefore contains implicitly the effect of superfluidity.
For finite system however, it is well known that effects like odd-even staggering requires the explicitly introduction of the anomalous 
density. The treatment of the pairing field with the new functional needs also to be clarified in the near future to extend the 
present work to finite systems. 

\begin{acknowledgments}
I would like to thank M. Grasso and J. Yang for numerous discussions that motivated this work 
and for their reading of the manuscript. I also thank G. Hupin for discussions at the early stage of this work and  
A. Gezerlis for his comments on the manuscript.
\end{acknowledgments}

\end{document}